\documentclass [prl,twocolumn,showpacs, superscriptsaddress]{revtex4}

\usepackage{footnote}
\usepackage[dvips]{graphicx}
\usepackage{color}
\usepackage{latexsym}
\usepackage{amsmath}
\usepackage{feynmf}

\newcommand{\ii}{\rm i}

\newcommand{\qvec}{\mathbf{q}}
\newcommand{\pvec}{\mathbf{p}}

\alph{footnote}

\begin{document}

\title{Many body exchange effects close to the s-wave Feshbach resonance in two-component Fermi systems: is a triplet superfluid possible?}
\author{Sergio~Gaudio}
\affiliation{ Dipartimento di Fisica, Universita' "La Sapienza" di Roma, 
  P.le Aldo
  Moro 2 and CNR-ISC, via Dei Taurini 19, Rome, 00185, Italy}
\author{Jason~Jackiewicz}
\affiliation{Max-Planck-Institut f\"{u}r Sonnensystemforschung, 37191 Katlenburg-Lindau, Germany}
\author{Kevin~S.~Bedell}
\affiliation{ Department of Physics, Boston College, 140 Commonwealth Ave, Chestnut
 Hill, MA, 02139, USA }

\begin{abstract}
We suggest that the exchange fluctuations close to a Feshbach resonance  in a
two-component Fermi gas can result in an effective p-wave attractive
interaction. On the BCS side of a Feshbach resonance, the magnitude of this
effective interaction is comparable to the s-wave interaction, therefore
leading to a possible spin-triplet superfluid in the range of temperatures of
actual experiments. We also show that the particle-hole exchange fluctuations introduce an
effective scattering length which does not diverge, as the standard mean-field one
does. Finally, using the effective interaction quantities we are able to model the molecular binding energy on the
BEC side of the resonance.
\end{abstract}

%\begin{keywords}
%Degenerate Fermi gas; Fermi liquid;  Feshbach resonance; superfluidity
%\end{keywords}\medskip
\maketitle
In the atomic Fermi gases such as $^{40}$K and $^6$Li, the use of Feshbach resonances has
opened the possibility of exploring the very interesting limit for which the
mean-field approximation predicts a smooth crossover from BEC to BCS pairing
as one goes through the resonance. At low energies, the inter-atomic
interaction is very well described by the s-wave scattering length,
$a_{\rm s}$. Moreover, no direct interactions are possible in the triplet channel.
In fact, higher-order expansions in the scattering length are suppressed at very
low temperatures and the symmetry of the wave function, due to Pauli
exclusion, does not allow s-wave scattering for fermionic atoms in the same
spin channel \cite{note1}.

Although the scattering length in the two-body problem is
diverging, it is instructive to consider the possibility of pairing
in the higher-order scattering channels due to exchange
fluctuations. It is also not clear whether atomic systems behave as
Fermi liquids (FL), or how similar they are with high $T_c$
superconductors (HTSC) or any other strongly correlated 
systems.

In this Letter, we want to show two things. Firstly, that it is possible to
build a Fermi liquid theory (FLT) in the atomic Fermi gases, particularly in the BCS region. This formalism  explains the basic features of
these gases like the scattering lengths, and possibly, the binding energies in
the strongly interacting regime, which is not accessible by simple
perturbation theory. Secondly, we show that important contributions can arise in
higher-order momentum channels. The resulting  triplet pairing
is comparable, in magnitude, to the s-wave one, and the correspondent triplet superfluid transition temperature is within experimental reach.
Since the triplet interaction can only occur through the fluctuations induced
by the strong interactions with the other spin channel, we focus our discussion primarily on the induced term. Its contribution may give an instability of the Fermi sea for  (quasi-)particles with equal
spins, and leads to a possible transition to a triplet superfluid. We will
keep the discussion quite general, since our approach may be of interest to
other fields. Then, we specify to the cold atom physics case, as we proceed.

In a Fermi liquid at sufficiently low temperatures ($T_{\rm c} < T \ll T_{\rm F}$, where $T_{\rm c}$
is the BCS transition temperature and $T_{\rm F}$ is the Fermi temperature),  
it was shown in \cite{Bedell} that it is possible to separate the Fermi
liquid parameters, $f_{\pvec \pvec'}$, which by construction do not contain
any zero sound terms \cite{agd}, into two sets of terms in the limit when
$p \sim p'$ (here, we assume the general notation $\mathrm q=(\qvec,\omega)$ and $q=|\qvec|$). One term is the direct interaction of the
quasiparticles (QP), and the other is the crossed term of the particle-hole
contribution. Dropping for the moment the spin indices, the Fermi liquid
parameters are
\begin{eqnarray}
-f_{\pvec\pvec'}\! &=& 2\pi \ii \, Z_{\pvec}\, Z_{\pvec'}
\lim_{\omega\rightarrow 0 }\lim_{\mathit{q}\rightarrow 0}\Gamma^{0}\left(p+\!\frac{q}{2},p-\frac{q}{2}; p\!-\!p'\right)\nonumber \\
\!\!\!\!\!\!&\!+&\!\sum_{\pvec_1,\pvec_2} \!\!f_{\pvec \pvec_1}\!\!\left(\frac{\delta n_{\pvec_1}}{\delta U_{\pvec_2}} \right) \!f_{\pvec_2 \pvec'},
\end{eqnarray}
where $Z_{\pvec_i}$ are the residues of the single particle Green's functions
at the pole of the QP, $n_{\pvec}$ is the Fermi distribution
function, $U$ is some interaction, and $\delta n_{\pvec_1}/\delta U_{\pvec_2}$
is related to the response function and can be obtained from the QP
transport equation \cite{FLT}. 
Restoring the spin indices, we denote the first term by $d^{\sigma\sigma'}_{\pvec\pvec'}$ and the second term by $I^{\sigma\sigma'}_{\pvec\pvec'}$.  The many-body effects in the QP interaction are therefore separated into two contributions:
\begin{equation}
\label{d-i}
f^{\sigma \sigma'}_{\pvec\pvec'} = d^{\sigma \sigma'}_{\pvec\pvec'} +I^{\sigma \sigma'}_{\pvec\pvec'} 
\end{equation}
where $d^{\sigma \sigma'}_{\pvec\pvec'}$, the direct term, includes only the
diagrams which are not particle-hole reducible and is equivalent to the T-matrix in the particle-particle channel. The induced term, $I^{\sigma \sigma'}_{\pvec\pvec'} $, has contributions from the exchange of virtual collective excitations among the quasiparticles, i.e. density, spin-density, current, spin-current fluctuations to name a few (see \cite{Bedell} for full
details). The implicit assumption, as it is for all Fermi liquid theories, is that all the relevant processes occur on the Fermi surface.

Consider, now,
any quantity, say $f_{\ell}^{\rm s,a}$. This is related to its counterpart
$f_{\pvec\pvec'}^{\sigma\sigma'}$ by the definition
\begin{equation}
F_{\pvec\pvec'}^{\sigma\sigma'}=N(0)f_{\pvec\pvec'}^{\sigma\sigma'}=\sum_\ell
(F_\ell^{\rm s} + \sigma\cdot\sigma' F_\ell^{\rm a})P_\ell(\hat{\pvec}\cdot\hat{\pvec}'),
\end{equation}
where $F_\ell^{\rm s,a}=N(0)f_\ell^{\rm s,a}$, $N(0)$ is the density of states
at the Fermi surface. The superscript $\rm {s \,(a)}$ indicates the symmetric
(anti-symmetric) contribution with respect to the spin, the subscript  $\ell$
indicates the Legendre component, and $P_\ell$ is the Legendre polynomial . Then, by expanding the Bethe-Salpeter equation for the QP interactions in
the limit $\omega/| \qvec | \rightarrow 0$ in a rotationally invariant
system into Legendre polynomials, it can be shown that \cite{FLT}
\begin{equation}
\label{AvsF}
A^{\rm s,a}_{\ell} =N(0)a^{\rm s,a}_{\ell} = \frac{F^{\rm s,a}_{\ell}}{1+F^{\rm s,a}_{\ell}/(2 \ell+1)}.
\end{equation}
Here,  $ A^{\rm s (a)}_{\ell} =N(0)a^{\rm s (a)}_{\ell}$ is the symmetric (anti-symmetric) Legendre components  of the
scattering amplitudes of the
quasiparticles. Note that these scattering amplitudes differ from the bare
scattering amplitudes, since they contain the many-body effects of the theory
through the QP interactions $f$.
\begin{figure}[!b]
 \includegraphics[ width=0.7\linewidth] {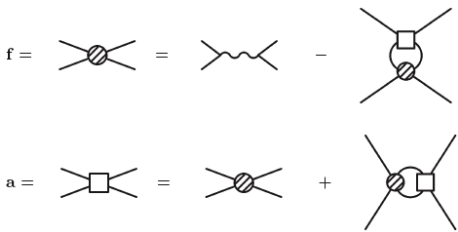}
\centering
        \caption{\label{LFT}Schematic diagrammatic relation between the Landau
          parameters and the scattering amplitudes defined in
          Eqs.~(\ref{d-i},\ref{AvsF},\ref{Fsinduced},\ref{Fainduced}).  The
          first term on the RHS of $f$ is the direct interaction $d$, and the
          second is the induced term $I$. Notice that when neglecting the induced terms,
          this description reduces to the RPA.}
\label{fig1}
\end{figure}
\noindent Given that Eq.~(\ref{AvsF}) is a non-perturbative result, it remains
valid even when $F_{\ell}$ diverges, since the $A_{\ell}$ remain finite. The only
approximation at  this point has been in assuming a Fermi liquid and the low
energy and momenta limits. From  \cite{Bedell},  it follows that
\begin{eqnarray}
\label{Fsinduced}
   F^{\rm s}_{\pvec\pvec'}& =  & D^{\rm s}_{\pvec\pvec'}+\frac{1}{2} \frac{F^{\rm s}_{0}\chi_{0}(q')F^{\rm s}_{0}}
   {1+F^{\rm s}_{0}\chi_{0}(q')}\!+\!
   \frac{3}{2}\frac{F^{\rm a}_{0}\chi_{0}(q')F^{\rm a}_{0}}
   {1+F^{\rm a}_{0}\chi_{0}(q')} ,\\
   \label{Fainduced}
F^{\rm a}_{\pvec\pvec'} &= &   D^{\rm a}_{\pvec\pvec'}+\frac{1}{2} \frac{F^{\rm s}_{0}\chi_{0}(q')F^{\rm s}_{0}}
   {1+F^{\rm s}_{0}\chi_{0}(q')}-
   \frac{1}{2}\frac{F^{\rm a}_{0}\chi_{0}(q')F^{\rm a}_{0}}
   {1+F^{\rm a}_{0}\chi_{0}(q')} ,
 \end{eqnarray}
where $q'^{2} = |\pvec-\pvec'|^2 = k^2_{\rm F}(1-\cos \theta_{\rm L})$ and
$\cos \theta_{\rm L} = \hat{\pvec}\cdot \hat\pvec'$ is the Landau angle, and $\chi_{0}(q')$ is the density-density correlation (Lindhard) functions (see \cite{Bedell}). Including $\ell\geq1$ terms is straightforward in this model, but only leads to small corrections to the results.   For the direct term $D$ in the low temperature
limit, the particle-particle T-matrix is proportional to the bare s-wave scattering
length $a_{\rm s}$. Since $D$ is then angle independent, it contributes only in the
$\ell=0$ momentum channel,  given by $D_{0}^{\rm s}=-D_{0}^{\rm a}=-N(0)U/2$, where $U =
4\pi\hbar^2a_{\rm s}/m $ is some on-site interaction. The direct interaction is antisymmetric and obeys
the Pauli principle $D_{0}^{\uparrow\uparrow(\downarrow\downarrow)}=0$. We
purposely neglect the remaining diagrams in the particle-particle channel,
since we are mostly interested in the induced interaction driven by the exchange of
collective excitations between the quasi-particles. In fact, at these temperatures,  this is the only
way a triplet interaction can arise in the same spin channel. We observe that
the form of the direct term depends on the model used, whereas the induced
term does not. Still, we emphasize that the resulting scattering lengths,
calculated through the Bethe-Salpeter equation, have the correct symmetries
and conserve the Pauli principle through the Landau sum rule, given by
$\sum_{\ell} (A^{\rm s}_{\ell}+A^{\rm a}_{\ell}) = 0$, which is not the case for the random phase
approximation (RPA).  Looking at the diagrams in Fig.~\ref{fig1} may help understand the
 differences. The RPA lack of the exchange terms in the particle-hole channel
 implies an inconsistent treatment of the QP interactions, since
the scattering amplitudes are not properly anti-symmetrized. The
 consequences of this will appear clear below. 

We now apply the above Fermi-liquid formalism to the specific case of the cold
atomic Fermi gases, in particular to  $^{40}$K. In these systems, the
scattering length can be varied by tuning the system close to a magnetic
Feshbach resonance\cite{pethick}. The s-wave scattering length $a_{\rm s}$
(denoted in the figures as ${\rm a_s^{bare}}$) varies as
\begin{equation}
\label{as}
a_{\rm s} = a_{\rm bg}\left( 1- \frac{\Delta B}{B-B_{0}}\right),
\end{equation}
where $B_{0}$ denotes the magnetic field value of the Feshbach resonance,
$\Delta B$ is the width of the resonance, and $a_{\rm bg}$ is the background
scattering length. Since most of the experimental systems deal with broad
resonances only, the contribution of the molecules from the closed channel can
be neglected \cite{hulet05}.
\begin{figure}[t!]
   \includegraphics[ width=0.8\linewidth]{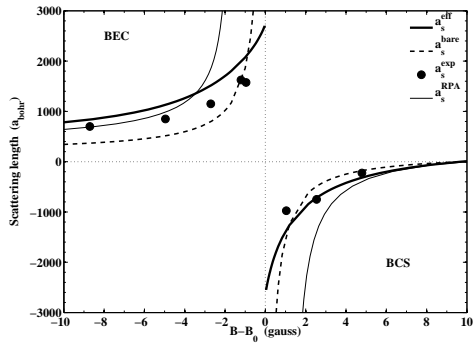}
\centering
   \caption{\label{fig:scattering_lengths}
S-wave scattering lengths (units of Bohr radii) as a function of the magnetic
field on both sides of the Feshbach resonance $B_0$, in $^{40}$K, using data from
\cite{regal2003}.  The particular scattering length is denoted in the
legend. The 50\% error bars in the experimental data have been removed
for clarity. We also plot the RPA scattering length for comparison.}
\end{figure}

We solve self-consistently Eqs.~(\ref{Fsinduced},\ref{Fainduced}) by varying the
 direct interaction $a_{\rm s}$  and use Eq.~(\ref{AvsF}) to obtain the scattering
 amplitudes $A_\ell^{\rm s,a}$ for $\ell=0, 1$. 
We then use these scattering amplitudes to  construct the singlet and triplet pairing amplitudes in the well-known
 s-p approximation \cite{FLT}, and call this the effective singlet scattering
 length $a^{\rm eff}_{\rm s}$.  In Fig.~\ref{fig:scattering_lengths} we show the
 results for the bare (Eq.~[\ref{as}], dashed line) and effective (thick line)
 s-wave
 scattering lengths calculated in this model for $^{40}$K on both sides of the
 Feshbach resonance $B=B_{0}$. The most important feature is that the
 effective scattering length does not diverge as the resonance is approached,
 while $a_{\rm s}$ does diverge.  Our results are very close in slope and magnitude to the
 experimental values.  On the other end, far from $B_{0}$, the effective and
 mean-field scattering lengths are comparable. We note that the presence of a
 strong p-wave interaction would influence the background scattering length
 $a_{\rm bg}$. However, since in that channel there is no resonance, the
 many-body effects will give a negligible contribution and we can safely
 assume the background scattering length to be constant.   

Also shown in  Fig.~\ref{fig:scattering_lengths} is the result predicted by
 RPA (thin line), which clearly fails to
 capture the correct physics as the resonance is approached. The divergence of
 the RPA scattering length implies the emergence of two new ground states on
 either side of the Feshbach resonance. On the BEC side, this would correspond
 to the Stoner instability or the onset of ferromagnetism. On the BCS side,
 this would correspond to phase separation. We note that neither of these two instabilities has been observed experimentally. 
\begin{figure}[b!]
   \includegraphics[ width=0.8\linewidth]{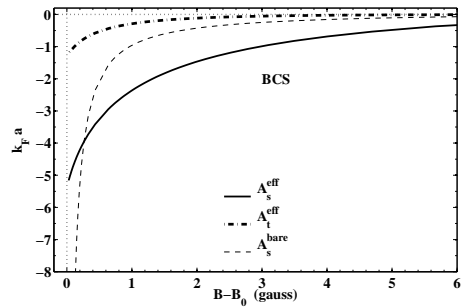}
\centering
  \caption{Scattering amplitudes on the (attractive) BCS side of the resonance $B_{0}$.
In our model, the effective (unitless) s-wave singlet $k_{\rm F}a_{\rm s}^{\rm
  eff}=A_{\rm s}^{\rm eff}$ and triplet $k_{\rm F}a_{\rm t}^{\rm
  eff}=A_{\rm t}^{\rm eff}$  amplitudes are finite and of similar magnitude at the
  resonance, while the bare interaction $k_{\rm F}a_{\rm s}=A_{\rm s}^{\rm bare}$  diverges, according to Eq.~(\ref{as}).}
\label{fig:scattering_amps}
\end{figure}

We now turn to the BCS side of the resonance where the scattering lengths are
negative, and
compare the effective triplet ($a_{\rm t}^{\rm eff}$) and effective singlet scattering lengths scaled
by the Fermi momentum $k_{\rm F}$. For magnetic fields near the resonance, the strength of the triplet and the
singlet potentials are actually comparable in magnitude, as shown in Fig.~\ref{fig:scattering_amps}.  
Therefore, although direct (triplet) pairing in the same hyperfine state is
suppressed initially at these low temperatures, the exchange of collective
excitations upon approaching the resonance drives a substantial attractive
interaction in the triplet channel through the induced interactions. Note that there is always attraction in
the triplet channel on both sides of the resonance, independent of the sign of the bare interaction
$U$. At low enough temperatures, it is not obvious that one can disregard
the possibility of having a triplet superfluid near resonance (on the BCS
side). In fact, below the triplet transition point, it seems quite reasonable
that these two many body states will compete, as
Fig.~\ref{fig:scattering_amps} suggests.

\begin{figure}[t]
   \includegraphics[ width=0.8\linewidth] {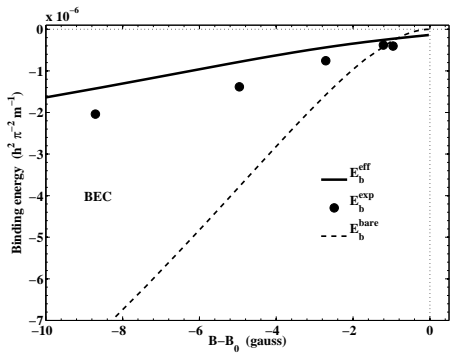}
\centering
   \caption{\label{fig:binding_energy}
   Binding energies on the BEC side in $^{40}$K, compared with data taken
   from\cite{regal2003}. The density used here is
   $n=5.8\times10^{13}$~cm$^{-3}$, with $\Delta B=9.7$~G and the Feshbach
   resonance, marked above with the vertical line, occurs at
   $B_{0}=224.21$~G.  $E_{\rm b}^{\rm eff}$ is the binding energy calculated from the
   effective scattering length in our theory, and $E_{\rm
   b}$ is the bare (mean-field) binding energy.}
\end{figure}
It is important  to compute the critical
temperatures expected for the various pairing instabilities. The expression is
very similar to the BCS one (see \cite{Anderson} and \cite{Leggett}).  The singlet
transition temperatures with our effective scattering amplitude could be as
large as $T^{\rm sing}_{c} \sim .7 \,T_{\rm F}$, while the triplet $T^{\rm trip}_{c} \sim .2\, T_{\rm F}$, if we use $T_{\rm F}$ as the cut-off scale. Note that these critical temperatures are quite high and that this is due to the use of the high-energy cut off. Also, there are numerous indications that singlet transition temperatures are of the order of  $.2T_{\rm F}$. This introduces a proportionality relation which gives a (upper bound of the) triplet transition temperature, within our framework, of $\sim .05 T_{\rm F}$. These temperatures are already obtainable in current experiments. 
We also mention that in the limit when the partial waves get very large, i.e., 
far away from the resonance, our approach gives the Gorkov and
Melik-Barkhudarov critical temperature\cite{GMB}, since the
particle-hole corrections become unimportant in this regime.

Lastly, we should remark that the present calculations hold for both equal
populations or for very small polarizations, $m=(n^\uparrow-n^\downarrow)/n$,
where $n^{\uparrow (\downarrow)}$ is the majority (minority) particle density,
and $n$ is the total density. For $m \ll n$, the corrections to the Fermi liquid parameters, which are quadratic in $m$, are, in fact, negligible. 
Recent experiments \cite{pol} have opened up the possibility of exploring the
triplet interactions of the system. Since the singlet BCS state is still
stronger than the triplet one, in order to see the triplet transition, it is
probably necessary to suppress the singlet superfluid. Indeed, even a small
polarizations, at low enough temperatures, might create the possibility of a
triplet superfluid in that channel. More likely, the presence of an external
field can establish a preferential direction and favor the triplet pairing,
similar to the A$_1$ phase in $^3$He. We emphasize, though, that the triplet
pairing is not only interesting in the superfluid phase, but also in the
normal one, as it will contribute to the properties of the system. In
fact, its thermodynamic properties, which we will discuss somewhere else, can be largely affected and therefore provide the experimental tools to verify the Fermi liquid behavior of the system close to resonance. 

Up to this point, we have assumed that the corrections due to the particle-particle contribution are not relevant.
In the normal phase, this assumption is plausible on the BCS side\cite{agd},
but on the BEC side can be justified only very close to the resonance and/or
for temperatures $k_{\rm B}T > E_{\rm b}$ (the experimental data are taken at
$T/T_{\rm F} \sim
.4$, although in this paper we are considering only the corrections due to the
quantum fluctuations $T\ll T_{\rm F}$ ), where $E_{\rm b}$ is the binding energy. Thus,  deep
into the BEC regime, our theory breaks down. On the other hand, we recover the
bare scattering length as soon as we get far away from the resonance and
therefore any FLT assumption is irrelevant. 

Thus, we might expect the theory to give a rough estimate of the scattering
lengths in the intermediate region as well. We therefore compute the binding
energy of the bound state in the open channel. It is calculated using the standard mean-field formula
$E_{\rm b}=-\hbar^{2}/ma_{\rm s}^{2}$, where $a_{\rm s}$ is the bare s-wave scattering
length. Since we lack a better estimate of the corrections to this formula due
to the many-body effects, we simply replace $a_{\rm s}$ with $a_{\rm s}^{\rm eff}$ and $m$
with $m^*=1+F_1^{\rm s}/3$, the effective mass in FL theory. The results are shown
in Fig.~\ref{fig:binding_energy}. The agreement with the experimental data is
quite surprising, but can be explained in terms of an effective Hamiltonian,
which, in the spirit of Landau's theory, progressively transforms the bare particles
into quasiparticles and the bare scattering length becomes the effective one
as the interaction increases. In this sense, one can adopt the same mean-field
formula of the binding energy, and indeed, the effective binding energy
profile reduces to the mean-field one in the weakly interacting regime. We
also note that the Fermi liquid parameters
represent only a part of the mean-field shift, since they do not contain the
zero sound channel contribution. The full contribution is instead given by the
full effective scattering length.
\begin{figure}[t]
   \includegraphics[ width=0.8\linewidth]{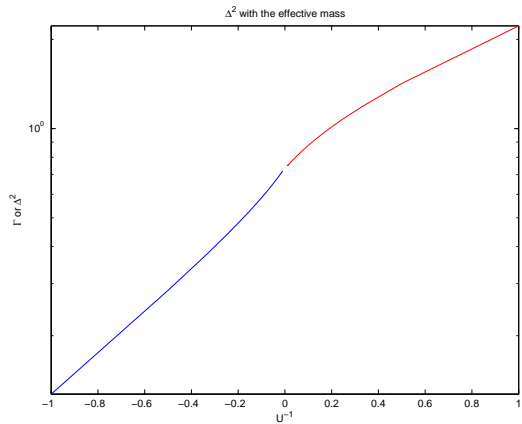}
\centering
  \caption{\label{fig:delta}
Gap squared using the weakly interacting limit equations as derived in \cite{Theory}, but replacing the effective mass and effective interaction from our theory to the bare ones. $U$ is the bare interaction strength as defined in the text. $\Gamma$ is related to the gap squared as in \cite{hulet05}, although our units do not correspond to those of the reference.}
\end{figure}

Engelbrecht~\textit{et al.}\cite{Theory} calculated the energy gap equation in
 the weakly interacting limit. They correctly pointed out that a comparison of
 the gap with the full solution should show roughly the extent of the strongly
 interacting regime. It is also interesting to note that the BEC-BCS crossover behavior is lost by using the weak-limit gap solution. In Fig.~(\ref{fig:delta}) we
 plot the energy gap corresponding to the weakly interacting BEC and BCS
 limits, but with our effective scattering lengths and masses. It is clearly
 seen that the crossover is re-established in terms of the
 quasiparticles. This shows that although the bare particles are strongly
 interacting, the quasiparticles may not, and hints that
 the gas in the normal phase is probably behaving as a Fermi liquid, even very
 close to the resonance.

In conclusion, we have built a theory which takes into account the
many body exchange effects in the quasiparticle-quasihole channel. This
theory, contrary to the RPA, respects the Pauli principle and
does not give spurious ground states. Inclusion of the exchange effects  is
therefore fundamental in obtaining the correct physics. We obtain a finite
scattering amplitude as seen experimentally. We have also shown that a triplet
superfluid is possible within the temperatures today achievable in cold atom
traps and that triplet paring should be taken into account when discussing the properties of the
system close to resonance. Furthermore, it seems possible in this formulation to
derive the basic properties
on the BEC side, although one should include properly the presence of the
bound  state, which we have not. The strong agreement with experiments
indicates that quasiparticles, not bare particles, are binding in the open channel. The good
interpolation of the intermediate interacting region between the BCS and BEC
sides is probably due to a
careful account of the particle-hole contributions in the theory. Finally, we remark  that
this approach, since it is not  restricted to the dilute gases, can be applied to other systems. 

It is our pleasure to thank Jan Engelbrecht, Eddy Timmermans,  J. Ho,
A. J. Leggett and R. Hulet for suggestions to the first version of the manuscript. One of
us, K.S.B, would like to thank the KITP and the Aspen Institute of Physics
for their hospitality.

\end{document}